\documentclass[a4paper]{article}

 \usepackage{graphicx,color}
 \usepackage{amsmath,amssymb}
\usepackage{times}
\usepackage{multirow}  
 \newtheorem{example}{Example}

\newcommand{\chv}{\ensuremath{\mathbf{v}}} 
\newcommand{\bfx}{\ensuremath{\mathbf{x}}} 
\newcommand{\bfp}{\ensuremath{\mathbf{p}}}

\newcommand{\bfc}{\ensuremath{\mathbf{c}}} 
\newcommand{\bfy}{\ensuremath{\mathbf{y}}} 

\newcommand{\bfX}{\ensuremath{\mathbf{X}}} 
\newcommand{\caL}{\ensuremath{\mathcal{L}}}

\newcommand{\bfO}{\ensuremath{\mathbf{O}}}
\newcommand{\bfe}{\ensuremath{\mathbf{e}}} 
\newcommand{\bfs}{\ensuremath{\mathbf{s}}} 

\newcommand{\pr}[1]{\ensuremath{\mathit{Pr}\!\left(#1\right)}} 

\newcommand{\lv}[1]{\mathbf{u}(#1)}
\newcommand{\tlv}[1]{\mathbf{\tilde u}(#1)}
\newcommand{\dlv}[2]{\mathbf{u}_{#1}(#2)}
\newcommand{\tdlv}[2]{\mathbf{\tilde u}_{#1}(#2)}
\newcommand{\zv}{\ensuremath{\mathbf{0}}}

\newcommand{\tsi}{\ell}

\newcommand{\algR}{AML}
\newcommand{\algS}{SVDL}
\newcommand{\algV}{SLA}
\newcommand{\algP}{PLA}

\begin{document}
 \date{}

\title{Parameter Identification for Markov Models of Biochemical Reactions}
\author{Aleksandr Andreychenko, Linar Mikeev, David Spieler, Verena Wolf\\[2ex]
 \begin{small}Saarland University,
         Saarbr\"ucken, Germany \end{small}}

 \maketitle
 
\begin{abstract}
 We propose a numerical technique for  parameter inference  
 in Markov models of biological processes. 
Based on time-series data of a process we estimate the kinetic 
rate constants by maximizing the likelihood of the data. 
The computation of the likelihood relies on a dynamic 
abstraction of the discrete state space of the Markov 
model which successfully mitigates the problem of state space 
largeness. 
 We compare two variants 
of our method to state-of-the-art, recently published 
methods  and demonstrate their usefulness and
efficiency on several case studies from systems biology.
\end{abstract}

\section{Introduction}
A widely-used  strategy in systems biology research is to refine
mathematical models of biological processes based on both computer simulations
and wet-lab experiments. 
In this context, parameter estimation methods for quantitative models play a
major role.
Typically, time series data is analyzed to learn the structure of a biochemical
reaction network and to calibrate the reaction rate parameters.
Direct measurement of parameters through wet-lab experiments is often difficult
or even impracticable.
There are extensive research efforts to estimate the reaction rate parameters
of ordinary differential equations (ODEs) that describe the evolution of the
chemical concentrations over time
(see, for instance, \cite{detinf1,detinf2,detinf4} and the references 
therein).
The problem of finding   parameters that minimize the difference between
observed  and predicted data is usually multimodal due to non-linear
constraints and thus requires global optimization techniques.

The assumption that chemical concentrations change deterministically and 
continuously in time is not always appropriate for biological processes.
In particular, if certain substances in the cell are present in small
concentrations the resulting stochastic effects cannot be adequately
described by deterministic models. In that case,  discrete-state stochastic
models are advantageous because they take into account the discrete 
  random nature of chemical reactions.
The theory of stochastic chemical kinetics provides a rigorously justified
framework for the description of chemical reactions where the effects of
molecular noise are taken into account~\cite{gillespie77}.
It is based on discrete-state Markov processes that explicitly represent
the reactions as state-transitions between population vectors. 
When the molecule numbers are large, the solution of the ODE description of a
reaction network and the mean of the corresponding stochastic model agree up to
a small approximation error. 
If, however, small populations are involved, then only
a stochastic description can provide probabilities of events of interest such as
probabilities of switching  between different expression states in gene
regulatory networks or the distribution of gene expression products.
Moreover, even the mean behavior of the stochastic model can largely deviate
from the behavior of the deterministic
model~\cite{loinger-lipshtat-balaban-biham07}. 
In such cases the parameters of the stochastic model rather then the
parameters of the deterministic model have to be
estimated~\cite{Tian2007,Timmer,Uz2010478}.

Here, we consider noisy time series measurements of the system state 
as they are available from wet-lab experiments. 
Recent experimental imaging
techniques  such as  high-resolution fluorescence microscopy can measure small
molecule counts with measurement errors of less than one
molecule~\cite{Golding}.
We assume that the structure of the underlying reaction network is known
but the rate parameters of the network are unknown.
Then we identify those
parameters that maximize the likelihood of the time series data.
Maximum likelihood estimators are the most popular estimators 
since they have desirable mathematical  properties. 
Specifically, they become minimum variance unbiased estimators 
and are asymptotically normal as the sample size increases. 

Our main contribution consists in devising an efficient algorithm for the 
numerical approximation of the likelihood and its derivatives w.r.t.
  the reaction rate constants. Previous techniques are based on 
Monte-Carlo sampling~\cite{Tian2007,Uz2010478} because the discrete state space 
of the underlying model is typically infinite in several dimensions and 
a priori a reasonable truncation of the state space is not availabe. 
Our method is not based on sampling but directly calculates the likelihood using
a dynamic truncation of the state space. More precisely, we first show that the
computation of the likelihood is equivalent to the evaluation of a product of
vectors and matrices. This product includes the transition probability matrix 
of the associated continuous-time Markov process, i.e., the solution of the
Kolmogorov differential equations (KDEs). Since solving the KDEs 
is infeasible, we propose two iterative approximation
algorithms during which the state space is
truncated in an on-the-fly fashion, that is, during a certain time interval we
consider only those  states that significantly contribute to the
likelihood. One approach exploits equidistant observation intervals
while the other approach is particularly well suited for observation intervals
that are not equidistant.
Both approaches take into account measurement noise during the observations.

 After introducing the  stochastic model in Section~\ref{sec:cme},
we discuss dynamic state space truncations for the transient probability 
distribution and its derivatives in Section~\ref{sec:approx}.
We introduce the maximum likelihood method in Section~\ref{sec:inference}
and present  the approximation methods in Section~\ref{sec:alg}.
Finally, we report on experimental results for two reaction networks 
(Section~\ref{sec:results})
and discuss related work in Section~\ref{sec:related}.

 \section{Discrete-state Stochastic Model}\label{sec:cme}
  According to Gillespie's theory of stochastic chemical kinetics, 
 a well-stirred mixture of $n$ molecular species in a volume 
 with fixed size and fixed temperature can be represented as 
 a continuous-time Markov chain 
 $\left\{\bfX(t),t\ge 0\right\}$~\cite{gillespie77}. 
 The random vector $\bfX(t)=\left(X_1(t),\ldots,X_n(t) \right)$ 
 describes the chemical populations at time $t$, i.e., $X_i(t)$ is 
 the   number of molecules of type $i\in\{1,\ldots,n\}$ at time 
$t$.
 Thus, the state space of $\bfX$ is $\mathbb Z^n_+=\{0,1,\ldots\}^n$.
The state changes of $\bfX$ are triggered by the occurrences of 
chemical reactions, which are of $m$ different types.  
For $j\in\{1,\ldots,m\}$  
let $\chv_j\in\mathbb Z^n$ be the nonzero
\emph{change vector} of the $j$-th reaction type, that is, 
$\chv_j=\chv_j^-+\chv_j^+$ where $\chv_j^-$ contains only 
non-positive entries, which specify how many molecules of each
species are consumed (\emph{reactants}) if an instance of the 
reaction occurs. 
The vector  $\chv_j^+$ contains only non-negative
 entries, which  specify how many molecules of each
species are produced (\emph{products}). 
Thus, if $\bfX(t)=\bfx$ for some $\bfx\in\mathbb Z^n_+$ with 
$\bfx+\chv_j^-$ being non-negative, then $\bfX(t+dt)=\bfx+\chv_j$
is the state of the system after the occurrence of the $j$-th 
reaction within the infinitesimal time interval $[t,t+dt)$.
 
Each reaction 
type has an associated \emph{propensity function}, denoted by 
$\alpha_1,\ldots,\alpha_m$, which is such that 
$\alpha_j(\bfx)\cdot dt$ is the probability that, given $\bfX(t)=\bfx$, 
one instance of the $j$-th reaction occurs within $[t,t+dt)$.
The value $\alpha_j(\bfx)$ is proportional to the number of distinct
reactant combinations in state $\bfx$. More precisely, if 
$\bfx=(x_1,\ldots,x_n)$ is a state for which $\bfx+\chv_j^-$ 
is nonnegative then, for reactions with at most two reactants,\vspace{-1ex}
\begin{equation}\label{eq:propf}
\textstyle\alpha_j(\bfx)=\left\{\!\!\begin{array}{l@{\,}c@{\,}l}
c_j & \mbox{ if } & \chv_j^-=(0,\ldots,0),\\
c_j \cdot x_i& \mbox{ if } & \chv_j^-=-\bfe_i,\\
c_j \cdot x_i\cdot x_\ell& \mbox{ if } & \chv_j^-=-\bfe_i-\bfe_\ell,\\
c_j \cdot {x_i\choose 2}=c_j \!\cdot\! \frac{x_i\cdot (x_i-1)}{2}& \mbox{ if } & \chv_j^-=-2\cdot \bfe_i,\\
\end{array}\right.\vspace{-1ex}
\end{equation}
where $ i\neq \ell$, $c_j>0$, and 
$\bfe_i$ is the vector with the $i$-th entry $1$ and all other
entries $0$.

\begin{example}
\label{ex:simplegene}
We consider the simple gene expression model  described in~\cite{Timmer}
that involves three chemical species, namely DNA$_{\text{ON}}$,
DNA$_{\text{OFF}}$, and mRNA, which are represented by the random variables
$X_1(t)$, $X_2(t)$, and $X_3(t)$, respectively. The three possible
reactions are 
DNA$_{\text{ON}} \to$ DNA$_{\text{OFF}}$, DNA$_{\text{OFF}} \to$ DNA$_{\text{ON}}$, and 
DNA$_{\text{ON}} \to$ DNA$_{\text{ON}} + $ mRNA. 
Thus, $\chv_1^- = (-1,0,0)$, $\chv_1^+ = (0,1,0)$, 
$\chv_2^- = (0,-1,0)$, $\chv_2^+ = (1,0,0)$,  $\chv_3^- = (-1,0,0)$
and $\chv_3^+ = (1,0,1)$. For a state $\bfx=(x_1,x_2,x_3)$,
the propensity functions are
$\alpha_1(\bfx) = c_1\cdot x_1$, $\alpha_2(\bfx) = c_2\cdot x_2$,
and $\alpha_3(\bfx) = c_3\cdot x_1$. Note that given the initial state 
$\bfx=(1,0,0)$, at any time, either the DNA is active or not, i.e.
$x_1 = 0$ and $x_2=1$, or $x_1=1$ and $x_2=0$. Moreover, 
the state space of the model is infinite in the third dimension.
For a fixed time instant $t>0$,    no upper bound on the number  
of mRNA is known a priori. 
All states $\bfx$ with $x_3 \in\mathbb Z_+$ 
have positive probability if $t>0$ but these probabilities will 
tend to zero as $x_3\to\infty$.
\end{example} 

In general, the reaction rate constants $c_j$ refer  to the 
probability that a randomly selected 
pair of reactants collides and undergoes the $j$-th chemical reaction. 
It depends on the volume and the temperature of the system as well
as on the microphysical properties of the reactant species. 
Since reactions of higher order (requiring 
more than two reactants) are usually the result of several successive lower 
order reactions, we do not consider the case of more than two reactants.


\textbf{The Chemical Master Equation.}
 For $\bfx\in \mathbb Z^n_+$ and $t\ge 0$, let $p(\bfx,t)$ denote the 
probability $\pr{\bfX(t)=\bfx}$ and let $\bfp(t)$ be the row vector 
with entries $p(\bfx,t)$.

Given $\chv_1^-,\ldots, \chv_m^-$, $\chv_1^+,\ldots, \chv_m^+$, 
$\alpha_1,\ldots,\alpha_m$, and some initial distribution $\bfp(0)$, the  
Markov chain $\bfX$ is uniquely specified
and its  evolution  is given by the 
 chemical master equation (CME)
 \begin{equation}
 \label{eq:CME}
 \begin{array}{r@{\ }c@{\ }l}
 \frac{d}{dt} \bfp(t) &=&  \bfp(t)Q,
  \end{array}
 \end{equation}
where  $Q$ 
is the infinitesimal generator matrix of $\bfX$
  with $Q(\bfx,\bfy)=\alpha_j(\bfx)$ if $\bfy=\bfx+\chv_j$
and $\bfx+\chv_j^-\ge 0$.
Note that, in
order to simplify our presentation, we assume here that all vectors $\chv_j$
are distinct. All remaining entries of $Q$ are zero except for the diagonal 
entries which are equal to the negative row sum. 
The ordinary first-order differential equation in~\eqref{eq:CME}  is a direct
consequence of the Kolmogorov forward equation.
Since $\bfX$ is a regular Markov process,~\eqref{eq:CME} 
has the general solution 
$
  {\bfp}(t)={\bfp}(0)\cdot e^{Qt},
$
 where $e^{A}$ is the matrix exponential of a matrix $A$.
If the state space of $X$ is infinite, 
then we can only compute approximations of  ${\bfp}(t)$.
But even if $Q$ is finite, its size is often large 
because it grows exponentially with the number of state variables. 
Therefore standard numerical solution techniques 
for  systems of  first-order linear equations of the form of~\eqref{eq:CME}
are infeasible.
The reason is that the number  of nonzero entries in $Q$ often 
exceeds the available 
 memory capacity  for systems of realistic size. 
 If  the populations of all species remain
 small (at most a few hundreds) 
 then  the CME can be efficiently
  approximated using projection methods~\cite{sliding,Munsky06,krylovSidje}
 or fast uniformization methods~\cite{FAUIET,Inexact}.
The idea of these methods is to avoid an exhaustive state space 
exploration and, depending on a certain time interval, restrict the analysis
of the system to a subset of states. 

Here, we are interested in the partial
derivatives of $\bfp(t)$ w.r.t. the 
reaction rate constants $\bfc=(c_1,\ldots,c_m)$. 
In order to explicitly indicate the
dependence of $\bfp(t)$ on the vector $\bfc$ we   write 
$\bfp(\bfc,t)$ instead of $\bfp(t)$ and $p(\bfx,\bfc,t)$ instead of 
$p(\bfx,t)$ if necessary.
We define the  row vectors $\bfs_j(\bfc,t)$   as the
derivative of $\bfp(\bfc,t)$ w.r.t. $c_j$, i.e.,
$$
\textstyle\bfs_j(\bfc,t)=\frac{\partial \bfp(\bfc,t)}{\partial c_j} =
\lim_{\Delta c\to 0}\frac{\bfp(\bfc+{\bf \Delta
c}_j,t)-\bfp(\bfc,t)}{\Delta c},
$$
where the vector ${\bf \Delta c}_j$ is zero everywhere except for the $j$-th
position that is equal to $\Delta c$. We denote the entry in $\bfs_j(\bfc,t)$
that corresponds to state $\bfx$ by $s_j(\bfx,\bfc,t)$.
Using~\eqref{eq:CME}, we find that $\bfs_j(\bfc,t)$ is the unique solution of 
the system of ODEs 
 \begin{equation}
 \label{eq:derivCME}
 \begin{array}{r@{\ }c@{\ }l}
 \frac{d}{dt} \bfs_j(\bfc,t) &=&  \bfs_j(\bfc,t)Q +
\bfp(\bfc,t)\frac{\partial}{\partial c_j}Q,
  \end{array}
 \end{equation}
where $j\in\{1,\ldots, m\}$.
The initial condition is $s_j(\bfx,\bfc,0)=0$ for all $\bfx$ and $\bfc$ 
since $p(\bfx,\bfc,0)$ is independent of $c_j$.  

 \section{Dynamic state space truncation}\label{sec:approx}
The parameter estimation method that we propose 
in Section~\ref{subsec:statebased}
builds on the approximation of the transient distribution 
$\bfp(t)$ and the derivatives $\bfs_j(\bfc,t)$ for all $j$ 
at a fixed time instant $t>0$. 
Therefore we now discuss how to
 solve~\eqref{eq:CME} and~\eqref{eq:derivCME}   simultaneously 
using an explicit fourth-order Runge-Kutta method and a dynamically 
truncated state space. 
This truncation is necessary because models of chemical 
reaction networks typically have a very 
large or infinite number of states $\bfx$ with 
nonzero values for $p(\bfx,t)$ and $s_j(\bfx,\bfc,t)$.
For instance, the system in Example~\ref{ex:simplegene}
is infinite in one dimension.
In order to keep the number of 
states, that are considered in a certain step of the numerical 
integration, manageable we suggest a dynamic 
 truncation of the state space that, for a given time 
interval, neglects those states being not relevant 
during that time, that is, we neglect states that 
have a probability that is smaller than a certain threshold.

 First, we remark that the equation that corresponds to 
state $\bfx$ in~\eqref{eq:CME} is given by
\begin{equation}\label{eq:CMEstate}
\textstyle \frac{d}{dt}p(\bfx,t)=\sum_{j:\bfx-\chv_j^-\ge 0} \alpha_j(\bfx-\chv_j) 
p(\bfx-\chv_j,t)-\alpha_j(\bfx)
p(\bfx,t).
\end{equation}
and it describes the  change of the probability of state $\bfx$ as
the difference between inflow of probability at rate 
$\alpha_j(\bfx-\chv_j)$ from direct predecessors $\bfx-\chv_j$
 and outflow of probability at rate $ \alpha_j(\bfx)$. 
Assume now that an initial distribution $\bfp(0)$ is given. 
We choose a small positive constant $\delta$ and define the 
set of significant states $S=\{\bfx\mid p(\bfx,0)>\delta\}$.
We then only integrate equations in~\eqref{eq:CME} 
and~\eqref{eq:derivCME} that belong to states in $S$.
If $h$ is the time step of the numerical integration,
then for the interval $[t,t+h)$ we use the following 
strategy to modify $S$ according to the probability flow.
We check for all successors $\bfx+\chv_j\not\in S$  of a state 
$\bfx \in S$ 
whether  $p(\bfx+\chv_j,t+h)$ becomes
   greater than $\delta$ at time $t+h$ as they receive ``inflow''
from their direct predecessors (see Eq.~\eqref{eq:CMEstate}).
 If the probability that $\bfx+\chv_j$ receives is greater $\delta$, 
then we add $\bfx+\chv_j$ to $S$. Note that since we use 
a fourth-order method, states reachable within at most four
transitions from a state in $S$ can be added during one step 
of the integration.  
On the other hand, whenever  $p(\bfx,t)$  
 becomes less or equal to $\delta$ for a state $\bfx\in S$ then we remove 
$\bfx$ from $S$. 
 We approximate  the probabilities and derivatives of all  
 states that are not considered during $[t,t+h)$ with zero. 
 In this way the computational costs of the numerical integration
 is drastically reduced, since typically 
 the number of states with probabilities 
less than $\delta$ is large and the main part of the 
probability mass is concentrated 
 on a small number of significant states. Due to
 the regular structure of $\bfX$, the probability of 
a state decreases exponentially
 with its distance to the ``high probability" locations. If $\delta$ is small 
 (e.g. $10^{-15}$) and the initial distribution is such that the main
 part of the probability mass (e.g. 99.99\%) distributes on
a manageable number of states,
 then even for long time horizons the approximation of the 
transient distribution is accurate.
For arbitrary Markov models, 
the approximation error of the derivatives could, in principle,
be large. For biochemical reaction networks, however, the 
underlying Markov process is well-structured and 
  the   sensitivity of the transient 
distribution w.r.t. the rate constants is comparatively 
small, i.e., small changes of the rate constants   result 
in a transient distribution that differs only slightly 
from the original distribution. 
Therefore, the derivatives of insignificant states are small and,
in order to calibrate parameters, it is sufficient to consider
the derivatives of probabilities of significant
states.
It is impossible to explore the whole 
state space and those parts containing most of the probability 
mass are most informative w.r.t. pertubations of the 
rate constants.
\begin{example}
\label{ex:enzyme}
We consider a simple enzyme reaction with three reactions that 
involve four different species,
namely enzymes (E), substrates (S), complex molecules (C), and 
proteins (P). The reactions are complex formation (E+S$\to$ C),
dissociation of the complex (C$\to$E+S), and protein production
(C$\to$E+P). The corresponding rate functions 
 are $\alpha_1(\bfx)=c_1\cdot x_1\cdot x_2$, 
$\alpha_2(\bfx)=c_2\cdot x_3$, and $\alpha_3(\bfx)=c_3\cdot x_3$
where $\bfx=(x_1,x_2,x_3,x_4)$.
The change vectors are given by $\chv_1^-=(-1,-1,0,0)$,
$\chv_1^+=(0,0,1,0)$, $\chv_2^-=(0,0,-1,0)$, $\chv_2^+=(1,1,0,0)$, 
$\chv_3^-=(0,0,-1,0)$, and $\chv_3^+=(1,0,0,1)$.
We start initially with probability one in state 
$\bfx=(1000,200,0,0)$ 
and compute $\bfp(t)$ and $\bfs_j(\bfc,t)$ for $t=10$,
$\bfc=(1,1,0.1)$, and $j\in\{1,2,3\}$. 
In Table~\ref{tab:enzyme} we list the results of the approximation 
of  $\bfp(t)$ and $\bfs_j(\bfc,t)$. We chose this model because it has 
a finite state space and we can compare our approximation with 
 the values obtained for $\delta=0$. The column ``Time'' lists 
the running times of the computation.
Obviously, the smaller $\delta$  
the more time consuming is the computation. 
The remaining columns refer to the maximum absolute error of all entries in the 
vectors $\bfp(t)$ and $\bfs_j(\bfc,t)$ where we use as exact values those 
obtained by setting $\delta=0$. Clearly, even if $\delta=0$ we have an 
approximation error due to the numerical integration of~\eqref{eq:CME} 
and~\eqref{eq:derivCME}, which is, however, very small compared to 
the error that originates from the truncation of the state space.
  \end{example}
 \begin{table}[t]
\begin{center}
\caption{Approximated transient distribution and 
  derivatives of the enzyme reaction network.\label{tab:enzyme}}
  \begin{tabular}{c@{\quad} c@{\quad} c@{\quad} c@{\quad} c@{\quad} c}
\hline   \hline 
\multirow{2}{*}{$\delta$} 	
& \multirow{2}{*}{Time} 
& \multicolumn{4}{c}{Maximum absolute error} \\
 &	& $\bfp(t)$	& $\bfs_1(\bfc,t)$ & $\bfs_2(\bfc,t)$ & $\bfs_3(\bfc,t)$\\	
\hline
0 & 10 h & 0 & 0 &0 & 0 \\
$10^{-20}$ & 47 sec & 1 $\cdot 10^{-11}$ & 1 $\cdot 10^{-12}$ & 1 $\cdot 10^{-12}$ & 4 $\cdot 10^{-9}$ \\
$10^{-15}$ & 25 sec & 1 $\cdot 10^{-11}$ & 8 $\cdot 10^{-11}$ & 9 $\cdot 10^{-11}$ & 2 $\cdot 10^{-8}$\\
$10^{-10}$ & 10 sec & 7 $\cdot 10^{-7}$ & 3 $\cdot 10^{-6}$   & 4 $\cdot 10^{-6}$  & 2 $\cdot 10^{-4}$ \\
\hline
  \end{tabular}
\end{center}\vspace{-3ex}
 \end{table}
A similar truncation effect can be obtained by sorting the 
entries of $\bfp(t)$ and successively  removing 
the smallest entries until a fixed amount $\varepsilon$ of probability 
mass is lost. If $\varepsilon$ is chosen proportional to the 
time step, 
then it is possible to bound the total approximation error of the 
probabilities, i.e., $\varepsilon=\tilde\varepsilon h/t$ where 
$\tilde\varepsilon$ is the total approximation error for a time horizon 
of length $t$. If memory requirements and running time 
 are more pressing then accuracy, then we can adjust 
the computational costs of the approximation 
 by keeping only the $k$ most probable  states in each step
for some integer $k$.

\section{Parameter Inference}\label{sec:inference}
Following the notation in~\cite{Timmer}, we assume that observations
of a biochemical network are made at time instances 
$t_1,\ldots,t_R\in\mathbb R_{\ge 0}$
where $t_1<\ldots<t_R$.
Moreover, we assume that $O_i(t_\ell)$ is the observed 
number of species $i$ at time 
$t_\ell$ for  $i\in\{1,\ldots,n\}$ and $\ell\in\{1,\ldots,R\}$. 
Let $\bfO(t_\ell)=\left(O_1(t_\ell),\ldots,O_n(t_\ell)\right)$ be
the corresponding vector of observations.
Since these observations are typically subject to measurement errors,
we assume that $O_i(t_\ell)=X_i(t_\ell)+\epsilon_i(t_\ell)$
where the error terms $\epsilon_i(t_\ell)$ are independent and 
identically normally 
distributed with mean zero and standard deviation $\sigma$. Note that 
$X_i(t_\ell)$ is the true population of the $i$-th species 
 at time $t_\ell$.
Clearly, this implies that, conditional on $X_i(t_\ell)$,
the random variable $O_i(t_\ell)$ is independent of all 
other observations as well as  independent of the history of $\bfX$ before 
time $t_\ell$.

We assume further that for the unobserved process $\bfX$
we do not know the values of the rate constants $c_1,\ldots,c_m$ and 
our aim is to estimate these constants.
Similarly, the exact standard deviation $\sigma$ of the error terms is 
unknown and must be estimated\footnote{We remark that it is straightforward 
to extend the estimation framework that we present in the sequel   such  
that a covariance matrix for a multivariate normal distribution 
of the error terms is estimated. In this way, different measurement errors  of
the species can be taken into account as well as dependencies between 
error terms. }.
Let $f$ denote the joint density of $\bfO(t_1),\ldots,\bfO(t_R)$.
Then the likelihood of the observations is~\cite{citeulike:821121}
\begin{equation}\label{eq:likelihood}
\begin{array}{lcl}
\caL&=&f\left(\bfO(t_1),\ldots,\bfO(t_R)\right)\\[1ex]
&=&\sum_{\bfx_1}\ldots \sum_{\bfx_R} f\left(\bfO(t_1),\ldots,\bfO(t_R)
\mid \bfX(t_1)=\bfx_1,\ldots,\bfX(t_R)=\bfx_R\right)\\[1ex]
&&   \pr{\bfX(t_1)=\bfx_1,\ldots,\bfX(t_R)=\bfx_R},
\end{array}
\end{equation}
that is, $\caL$ is the probability to observe 
$\bfO(t_1),\ldots,\bfO(t_R)$. Note that $\caL$ depends 
on the chosen rate parameters $\bfc$ since the probability 
measure $ \pr{\cdot }$ does. Furthermore, $\caL$ depends on 
$\sigma$ since the density $f$ does.
 When necessary, we will make this 
dependence explicit by writing $\caL(\bfc,\sigma)$ instead of $\caL$.
We now seek constants $\bfc^*$ and a standard deviation $\sigma^*$ such that 
\begin{equation}\label{eq:MLEestimator}
 { \caL(\bfc^*,\sigma^*)=\max_{\sigma,\bfc} \caL(\bfc,\sigma)}
\end{equation}
where  the maximum is taken over all $\sigma>0$ and vectors
$\bfc$ with all components strictly positive. 
This optimization problem is known as the maximum likelihood 
problem~\cite{citeulike:821121}.
Note that $\bfc^*$ and $\sigma^*$ are random variables because 
they depend on the (random) observations $\bfO(t_1),\ldots,\bfO(t_R)$.

If more than one sequence of observations is made, then
the corresponding likelihood is the product of the likelihoods 
of all individual sequences.  More precisely, if 
$\bfO^k(t_l)$ is the $k$-th observation that 
has been observed  at time instant $t_l$ 
where $k\in\{1,\ldots,K\}$, then we define 
$\caL_k(\bfc,\sigma)$ as the probability to observe 
$\bfO^k(t_1),\ldots,\bfO^k(t_R)$ and maximize 
\begin{equation}\label{eq:prodL}
 \textstyle\prod_{k=1}^K \caL_k(\bfc,\sigma). 
\end{equation}
In the sequel, we concentrate on expressions for
$\caL_k(\bfc,\sigma)$ and 
$\frac{\partial }{\partial c_j} \caL_k(\bfc,\sigma)$.
We first assume $K=1$ and drop index $k$. 
We consider the case $K>1$ later.
In~\eqref{eq:likelihood} we sum over all state  sequences
$\bfx_1,\ldots,\bfx_R$ such that 
$\pr{\bfX(t_\ell)=\bfx_\ell,1\le \ell\le R}>0$. 
Since $\bfX$ has a large or even infinite state space,
it is computationally infeasible to explore all possible 
sequences. 
In Section~\ref{sec:alg} we propose an algorithm 
to approximate the likelihoods and their derivatives.
We truncate the state space in
a similar way as in Section~\ref{sec:approx} and use
 the fact that~\eqref{eq:likelihood} can be written
as a product of vectors and matrices.  
Let $\phi_\sigma$ be the density of the normal distribution with 
mean zero and standard deviation $\sigma$.
Then
$$
\begin{array}{lcl}
&& f\left(\bfO(t_1),\ldots,\bfO(t_R)
\mid \bfX(t_1)=\bfx_1,\ldots,\bfX(t_R)=\bfx_R\right)\\[1ex]
&=&\prod_{\ell=1}^R
\prod_{i=1}^n f\left(O_i(t_\ell)\mid X_i(t_\ell)=x_{i\ell}\right)\\[1ex]
&=& \prod_{\ell=1}^R
\prod_{i=1}^n \phi_\sigma(O_i(t_\ell)-x_{i\ell}),
\end{array}
$$ 
where $\bfx_\ell=(x_{1\ell},\ldots,x_{n\ell})$.
If we write $w(\bfx_\ell)$ for   
$\prod_{i=1}^n \phi_\sigma(O_i(t_\ell)-x_{i\ell})$,
then the sequence $\bfx_1,\ldots,\bfx_R$ has weight 
$\prod_{\ell=1}^R w(\bfx_\ell)$ and, thus, 
\begin{equation}\label{eq:sumoverpaths}
\caL=\sum_{\bfx_1}\ldots \sum_{\bfx_R}  
   \pr{\bfX(t_1)=\bfx_1,\ldots,\bfX(t_R)=\bfx_R} \prod_{\ell=1}^R w(\bfx_\ell).
\end{equation}
Moreover, for the probability of the sequence $\bfx_1,\ldots,\bfx_R$ 
we have 
$$
\pr{\bfX(t_1)=\bfx_1,\ldots,\bfX(t_R)=\bfx_R}=p(\bfx_1,t_1) 
P_2(\bfx_1,\bfx_2)\cdots P_{R}(\bfx_{R-1},\bfx_R)
$$
where $P_\ell(\bfx,\bfy)=\pr{\bfX(t_{\ell})=\bfy\mid\bfX(t_{\ell-1})=\bfx}$.
Hence,~\eqref{eq:sumoverpaths} can be written as
\begin{equation}\label{eq:sumoverpaths2}
\caL=\sum_{\bfx_1}p(\bfx_1,t_1) w(\bfx_1)\sum_{\bfx_2}P_2(\bfx_1,\bfx_2)
 w(\bfx_2)  \ldots \sum_{\bfx_R}  P_{R}(\bfx_{R-1},\bfx_R) w(\bfx_R).
 \end{equation}
Let $P_\ell$ be the matrix with entries $P_\ell(\bfx,\bfy)$ for all states 
$\bfx,\bfy$. Note that  $P_{\ell}$ is the transition 
probability matrix of $\bfX$ for time step $t_{\ell}-t_{\ell-1}$
and thus the
general solution $e^{Q(t_{\ell}-t_{\ell-1})}$ of the 
Kolmogorov forward and backward differential equations
$$ \textstyle 
\frac{d}{dt}P_{\ell}= Q P_{\ell}, \hspace{10ex}
\frac{d}{dt}P_{\ell}=P_{\ell} Q.
$$
Using  
$\bfp(t_1)=\bfp(t_0)P_1$ with $t_0=0$, we can write~\eqref{eq:sumoverpaths2}
  in matrix-vector form as
\begin{equation}
\label{eq:likelihood_vm}
 \caL= \bfp(t_0) P_1 W_1 P_2 W_2 \cdots P_{R} W_R \bfe.
\end{equation}
Here,  $\bfe$  is the vector with all entries equal to one and
$W_\ell$ is a diagonal matrix whose diagonal entries are all equal to
$w(\bfx_\ell)$ with $\ell\in\{1,\ldots,R\}$, where $W_\ell$ is of the
same size as $P_\ell$. 
Since it is in general not possible to analytically obtain parameters that
maximize $\caL$, we use   optimization techniques to find $\bfc^*$ 
and $\sigma^*$. Typically, such techniques iterate over   values
of $\bfc$ and $\sigma$ and increase the likelihood $\caL(\bfc,\sigma)$
by following the gradient. 
Therefore, we need to   calculate the 
derivatives $\frac{\partial }{\partial c_j} \caL$ and $\frac{\partial }{\partial
\sigma} \caL$.  
For  $\frac{\partial }{\partial c_j} \caL$ we obtain 
\begin{equation}
\label{eq:likelihood_derivative_vm}
	\begin{array}{lcl}
	\frac{\partial }{\partial c_j} \caL &=& 
	\frac{\partial }{\partial c_j} \left( \bfp(t_0) P_1 W_1 P_2 W_2 \cdots
  P_{R} W_R \bfe \right)\\[1ex]
	 &=& \bfp(t_0) \left(\sum_{\ell=1}^R  \left(\frac{\partial }{\partial
c_j} P_{\ell} \right) W_{\ell}\prod_{\ell'\neq \ell} P_{\ell'} W_{\ell'} \right)
\bfe.
 	\end{array}
\end{equation}
The derivative of $\caL$ w.r.t. the standard deviation $\sigma$ is derived
 analogously. The only difference is that $P_1,\ldots,P_{R}$ are independent 
of $\sigma$ but $W_1,\ldots, W_R$ depend on $\sigma$.
 It is also important to note that   expressions for partial derivatives of
second order can be derived in a similar way. These derivatives can then be used
for an efficient gradient-based local optimization.
  
For $K>1$ observation sequences we can   
maximize the log-likelihood
\begin{equation}
\label{eq:logProdL}\textstyle
\log \prod_{k=1}^K \caL_k = \sum_{k=1}^K \log \caL_k,
\end{equation}
instead of the likelihood in~\eqref{eq:prodL},
where we abbreviate $\caL_k(\bfc,\sigma)$ by $\caL_k$.
Note that the derivatives are then given by
\begin{equation}
\label{eq:logDeriv}\textstyle
\frac{\partial}{\partial \lambda} \sum_{k=1}^K \log \caL_k = 
\sum_{k=1}^K \frac{\frac{\partial}{\partial \lambda} \caL_k}{\caL_k},
\end{equation}
where $\lambda$ is either $c_j$ or $\sigma$.
 It is also important to note that only the weights $w(\bfx_\ell)$
depend on $k$, that is, on the observed sequence
$\bfO^k(t_1),\ldots,\bfO^k(t_R)$. Thus, when we compute  
$\caL_k$ based on \eqref{eq:likelihood_vm} we use for all $k$ the same 
transition matrices $P_1,\ldots,P_{R}$ and the same initial 
conditions $\bfp(t_0)$, but possibly different  matrices $W_1,\ldots,W_R$.

\section{Numerical approximation algorithm} \label{sec:alg}
In this section, we focus on the numerical approximation of the 
likelihood 
and the corresponding 
derivatives w.r.t. the rate constants $c_1,\ldots,c_m$.
We propose two approximation algorithms for the likelihood and its 
derivatives, a state-based likelihood approximation (\algV) 
and a path-based likelihood approximation (\algP).
Both are based  on a 
dynamic truncation of the state space as suggested in Section~\ref{sec:approx}.
They differ in that the \algP~method exploits equidistant time series,
that is, it is particularly efficient if $h=t_{\ell+1}-t_\ell$ for all 
$\ell$ and if $\sigma$ is not too large. 
The \algV~algorithm   works for arbitrarily spaced  
time series and is efficient even if $\sigma$ is large.

 \subsection{State-based likelihood approximation}\vspace{-0.8ex}
 \label{subsec:statebased}
The \algV~algorithm calculates an approximation of the likelihood 
based on~\eqref{eq:likelihood_vm} by traversing the matrix-vector product 
from the left to the right. The main idea behind the algorithm is that 
instead of explicitly computing the matrices $P_\ell$, we express 
the vector-matrix product $\lv{t_{\ell-1}} P_\ell$ as a system of 
ODEs similar to the CME (cf. Eq.~\eqref{eq:CME}). 
Here, $\lv{t_0}, \ldots,\lv{t_R}$ are row vectors
obtained during the iteration over time points $t_0, \ldots,t_R$,
that is, we define
$\caL$ recursively as $\caL = \lv{t_R} \bfe$ with $\lv{t_0}  =  \bfp(t_0)$
and 
$$ 
\begin{array}{rcl}
  \lv{t_\tsi} & = & \lv{t_{\tsi-1}} P_{\tsi} W_{\tsi} \qquad \text{ for all } 1 \leq
\tsi \leq R,
\end{array}
$$
where $t_0=0$. Instead of computing $P_{\tsi}$ explicitly, we solve $R$ 
systems of ODEs
\begin{equation}\label{eq:ODEL}
\textstyle\frac{d}{dt} \tlv{t}=  \tlv{t} Q 
\end{equation}
with initial condition  $\tlv{t_{\ell-1}}=\lv{t_{\ell-1}}$ for the time
interval $[t_{\ell-1},t_\ell)$ where $\ell\in\{1,\ldots,R\}$. 
After solving the $\ell$-th system of ODEs we  set
$\lv{t_{\ell}}=\tlv{t_{\ell}}W_\ell$ and  finally  compute 
$\caL=\lv{t_R}\bfe$.
Since this is the same as solving the CME for different initial conditions, 
  we can use the   dynamic truncation of the state 
space proposed in Section~\ref{sec:approx}. 
Since the vectors $\tlv{t_{\ell}}$ do not 
sum up to one, we scale all entries by 
multiplication with $1/(\tlv{t_{\ell}}\bfe)$. 
This simplifies the truncation of the state space using the  
significance threshold $\delta$ since 
after scaling it can be interpreted as a probability.
In order to obtain the correct (unscaled) 
likelihood, we   compute $\caL$ as 
$\caL=\prod_{\ell=1}^R(\tlv{t_{\ell}}\bfe)$. 
We handle the derivatives of $\caL$ in a similar way. To shorten  our 
presentation, we only consider the derivative $\frac{\partial }{\partial c_j}
\caL$ in the sequel. An iterative scheme for $\frac{\partial }{\partial \sigma}
\caL$ is derived analogously.
From~\eqref{eq:likelihood_derivative_vm} we obtain 
$\frac{\partial }{\partial c_j} \caL = \dlv{j}{t_R}\bfe$ with
$\dlv{j}{t_0}  =  \zv \nonumber$ and
$$
\begin{array}{rcl}
  \dlv{j}{t_\tsi} & = & (\dlv{j}{t_{\tsi-1}} P_{\tsi} + \lv{t_{\tsi-1}}
\frac{\partial }{\partial c_j} P_{\tsi}) W_{\tsi} \qquad \text{ for all } 1
\leq \tsi \leq R,
  \label{eq:dlv_rec}
\end{array}
$$
where $\zv \nonumber$ is the vector with all entries zero.
Thus, during the solution of the $\ell$-th ODE in \eqref{eq:ODEL} 
we simultaneously solve
\begin{equation}\label{eq:ODEdL}
\textstyle\frac{d}{dt} \tdlv{j}{t}=  \tdlv{j}{t} Q + \tlv{t} \frac{\partial}{\partial c_j}Q
\end{equation}
with initial condition  $\tdlv{j}{t_{\ell-1}}=\dlv{j}{t_{\ell-1}}$ for the time
interval $[t_{\ell-1},t_\ell)$. As above,  we set
$\dlv{j}{t_{\ell}}=\tdlv{j}{t_{\ell}}W_\ell$ and 
  obtain $\frac{\partial }{\partial c_j}
\caL$ as $\dlv{j}{t_{R}}\bfe$.

Solving \eqref{eq:ODEL} and
\eqref{eq:ODEdL} simultaneously is equivalent to the computation 
of the partial derivatives in~\eqref{eq:derivCME} with different 
initial conditions. Thus,  we can   use   the approximation
algorithm proposed in Section~\ref{sec:approx} to 
approximate   $\dlv{j}{t_{\ell}}$. 
Experimental results  of the finite enzyme reaction network 
(see Example~\ref{ex:enzyme}) 
 show that the approximation errors 
of the likelihood and its derivatives are of the same order 
of magnitude as those  of the transient probabilities 
and their derivatives (not shown). 
 Note, however, that, if $\sigma$ is small 
only few states contribute significantly to the likelihood. 
In this case, truncation strategies based on sorting 
of vectors are more efficient without considerable accuracy losses
since the main part of the likelihood concentrates on very few entries
(namely those that correspond to states that 
are close to the observed populations).
 
In the case of $K$ observation sequences we repeat the  above algorithm 
   in order to sequentially compute $\caL_k$ for $k\in\{1,\ldots,K\}$.
We exploit~\eqref{eq:logProdL} and~\eqref{eq:logDeriv}
to compute the total log-likelihood and its derivatives as a sum of 
individual terms. 
Obviously, it is possible to parallelize the \algV~algorithm by 
computing    $\caL_k$ in parallel for all $k$.

\subsection{Path-based likelihood approximation}\vspace{-0.8ex}
If $\Delta t=t_{\ell}-t_{\ell-1}$ for all 
$\ell$ then the matrices $P_1,\ldots,P_{R}$ in \eqref{eq:likelihood_vm} are
 equal to 
the $\Delta t$-step transition matrix $T(\Delta t)$ with entries 
$\pr{\bfX(t+\Delta t)=\bfy\mid \bfX(t)=\bfx}$. Note that 
since we consider a time-homogeneous Markov process $\bfX$,
the matrix $T(\Delta t)$ is independent of $t$. 
The main idea of 
the \algP~method is to iteratively compute those parts of  $T(\Delta t)$ 
that correspond to state sequences (paths) $\bfx_1,\ldots,\bfx_R$
that contribute significantly  to $\caL$. The algorithm 
can be summarized as follows, where we omit the argument $\Delta t$ of $T$ 
to improve the readability and refer to the entries of $T$ as $T(\bfx,\bfy)$:
\begin{enumerate}
 \item We compute the transient distribution $\bfp(t_1)$ and its derivatives 
(w.r.t. $\bfc$ and $\sigma$)
as outlined in Section~\ref{sec:approx} using a significance threshold $\delta$.
\item For each state $\bfx_1$ with significant probability $p(\bfx_1,t_1)$ 
we approximate the rows of $T$ and $\frac{\partial}{\partial c_j}T$
that correspond to $\bfx_1$ based on a 
transient analysis for $\Delta t$ time units. 
More precisely, if $\bfe_{\bfx_1}$ is the vector with 
all entries zero except for the entry that corresponds 
to state $\bfx_1$ which is one, then we 
 solve~\eqref{eq:CME} with initial condition $\bfe_{\bfx_1}$
for $\Delta t$ time units in order to approximate  
$T(\bfx_1,\bfx_2)$ and $\frac{\partial}{\partial c_j}T(\bfx_1,\bfx_2)$ 
for all $\bfx_2$.
During this  transient analysis we again
apply   the dynamic truncation of the state space 
proposed in Section~\ref{sec:approx} with threshold $\delta$.
\item We then store for each  pair $(\bfx_1,\bfx_2)$  
 the (partial) likelihood $a(\bfx_1,\bfx_2)$  
and its derivatives:
 $$\begin{array}{rcl}
a(\bfx_1,\bfx_2)&=& p(\bfx_1,t_1)\cdot w(\bfx_1)\cdot T(\bfx_1,\bfx_2) \cdot w(\bfx_2)\\[1ex]
\frac{\partial}{\partial c_j}a(\bfx_1,\bfx_2)&=&
\frac{\partial}{\partial c_j} p(\bfx_1,t_1)\cdot w(\bfx_1)\cdot T(\bfx_1,\bfx_2) \cdot w(\bfx_2) \\  && +
p(\bfx_1,t_1)\cdot w(\bfx_1)\cdot \frac{\partial}{\partial c_j} T(\bfx_1,\bfx_2) \cdot w(\bfx_2).
\end{array}
$$
 \item We    reduce the number of considered pairs  
by sorting  $a(\bfx_1,\bfx_2)$ for all pairs  $(\bfx_1,\bfx_2)$
calculated in the previous step  
 and keep the most probable pairs (see also Section~\ref{sec:approx}).
 \item Next, we repeat steps 2-4, where in step 2 we start the analysis 
from all states $\bfx_2$ that are the last element of 
a pair kept in the previous step. In step 3 we
store triples of states, say, $(\bfx_1,\bfx_2,\bfx_3)$ 
and recursively compute their likelihood and the corresponding 
derivatives by multiplication
with  $T(\bfx_2,\bfx_3)$ and   $w(\bfx_3)$, i.e., for the likelihood we compute
 $$\begin{array}{rcl}
a(\bfx_1,\bfx_2,\bfx_3)&=&a(\bfx_1,\bfx_2)\cdot T(\bfx_2,\bfx_3) \cdot w(\bfx_3)\\[1ex]
\frac{\partial}{\partial c_j}a(\bfx_1,\bfx_2,\bfx_3)&=&\frac{\partial}{\partial
c_j}a(\bfx_1,\bfx_2)\cdot T(\bfx_2,\bfx_3) \cdot w(\bfx_3)\\
&&
+a(\bfx_1,\bfx_2)\cdot
\frac{\partial}{\partial c_j}T(\bfx_2,\bfx_3) \cdot w(\bfx_3).
\end{array}
$$
Note that we may reuse some of the entries of $T$ since they 
already have been calculated in a previous step. In step 4 we
again reduce the number  of triples  $(\bfx_1,\bfx_2,\bfx_3)$   by 
sorting them  according to their likelihood. We then keep the 
  most probable triples, and so on.
Note that in step 4 we cannot use a fixed truncation threshold 
$\delta$ to reduce the number of state sequences (or paths) 
since their probabilities may become very small as the sequences become longer.
\item We stop the prolongation of paths $\bfx_1,\ldots,\bfx_\ell$ 
when the   
time instance $t_R=\Delta t\cdot R$ is reached and compute 
an approximation of $\caL$ and its derivatives by summing up 
the corresponding values of all paths
(cf. Eq.~\eqref{eq:sumoverpaths}).
\end{enumerate}
If we have more than one observation sequence, i.e., 
$K>1$, then we repeat the procedure to compute 
$\caL_k$ for all $k$ and use~\eqref{eq:logProdL} to calculate the 
total log-likelihood. 
 Note that the contribution of each path
$\bfx_1,\ldots,\bfx_R$ to $\caL_k$ may be different for each $k$. 
It is, however, likely that the entries of $T$ can be reused not only during 
the computation of each single $\caL_k$ but also for different values of $k$.
If many entries of $T$ are reused during the computation, the algorithm
performs fast compared to other approaches.
For our experimental results in Section~\ref{sec:results}, we
keep the ten most probable paths in step 4. 
Even though this enforces  a coarse approximation, the 
likelihood is approximated very accurately if $\sigma$ is small, since in this 
case only few  paths contribute significantly to $\caL_k$. On the other
hand, if $\sigma$ is large, then the approximation may become inaccurate
depending on the chosen truncation strategy.  
Another disadvantage of the \algP~method is that for 
non-equidistant time series,
the performance is slow since we have to compute (parts of) different 
transition matrices and, during the computation of $\caL_k$, the transition 
probabilities cannot be reused.

\section{Experimental results}\label{sec:results}
  In this section we present experimental results of the  \algV~and
   \algP~method. For equidistant time series, we compare 
our approach to the approximate maximum likelihood (AML)  
and the singular value decomposition   (SVDL) method described by Reinker
et al.~\cite{Timmer} (compare also Section~\ref{sec:related}). Since an
implementation of
the AML and SVDL method
  was not available to us, we chose the same examples and experimental 
conditions for the time series as Reinker et al. and compared our results to 
those listed in the results section in~\cite{Timmer}.
We also consider non-equidistant time series. To the best of our knowledge there
exists no direct numerical approach for non-equidistant time series with
measurement error that is based on the maximum likelihood method.

We generated time series data for two different examples from systems biology 
 using Monte-Carlo simulation~\cite{gillespie77} and added  error terms 
$\epsilon_i(t_\ell)$ to the population of the $i$-th species at time $t_\ell$.
Besides the   simple network 
described in Example~\ref{ex:simplegene} we consider a 
  more complex network with eight reactions
and five   species for  the transcription regulation of a
repressor protein~\cite{Timmer}:
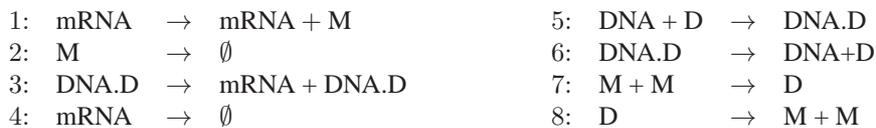
\begin{center}
$ \begin{array}{r@{: \quad }lcl@{\hspace{12ex}}r@{: \quad }lcl}
1 &  \mbox{mRNA}  & \to & \mbox{mRNA} + \mbox{M} & 5 &  \mbox{DNA + D} & \to &
\mbox{DNA.D}\\
2 &  \mbox{M}     & \to & \emptyset              & 6 &  \mbox{DNA.D}   & \to &
\mbox{DNA+D}\\
3 &  \mbox{DNA.D} & \to & \mbox{mRNA + DNA.D}    & 7 &  \mbox{M + M}   & \to &
\mbox{D}\\
4 &  \mbox{mRNA}  & \to & \emptyset              & 8 &  \mbox{D}       & \to &
\mbox{M + M}\\
  \end{array}
  $
\end{center}
The initial molecular populations are $(2,4,2,0,0)$ for
M, D, DNA, mRNA, and DNA.D.
The reachable state space of the model is infinite
in three dimensions since the populations of mRNA, M, and D are  unbounded.
The rate constants are  
$\bfc = (0.043, 0.0007,$ $0.0715, 0.00395, 0.02,0.4791, 0.083,
0.5)$.
For the network in Example~\ref{ex:simplegene} we chose the same 
parameters as Reinker et al., namely  $\bfc = (0.0270, 0.1667, 0.40)$.
  
 For the generation of time series data we fix the (true)  constants
$\bfc$  and the standard deviation $\sigma$ of the error terms. 
We use the SLA and PLA method
to estimate $\bfc$ and $\sigma$ such that   the likelihood of the
time series becomes maximal under these parameters. 
Since in practice only few observation sequences are available, we 
estimate the parameters based on $K=5$ observation
sequences. As suggested by Reinker et al., we repeat the  generation of batches
of five observation sequences and the estimation of parameters 100 times
to approximate the mean and the standard deviation of the estimators.

Our algorithms for the approximation of the likelihood
are implemented in C++ and we run them  on an Intel Core
i7 at 2.8 Ghz with 8 GB main memory. They are linked to MATLAB's
optimization toolbox which we use to minimize the negative log-likelihood.
Since we use a global optimization method (MATLAB's
global search),   the running time of our method depends on the 
tightness of the  intervals that we use as constraints for the unknown 
parameters
 as well as on the number of starting points of the
global search procedure.  We chose intervals that correspond to the  
order of magnitude of the parameters, i.e., if $c_j\in O(10^n)$ for some 
   $n\in\mathbb Z$ then we use 
the interval $[10^{n-1},10^{n+1}]$ as constraint for $c_j$. 
E.g. if $c_j=0.1$ then $n=-1$ and we use the interval
 $[10^{-2},10^0]$. Moreover, for global search we used 20 starting points for
the gene expression example and 5 for the transcription regulation example. 
Note that this is the
only difference of our experimental conditions compared to Reinker et al. who
use a local optimization method and start the optimization with the true
 parameters.

In both algorithms we choose a significance 
threshold of $\delta=10^{-15}$.  
Since the \algP~method becomes slow if the number of paths that 
are considered is large, in step 4 of the algorithm we
 reduce the number of paths that we consider by keeping 
only the 10 most probable paths. In this way, the computational 
effort of the \algP~method remains tractable even in the 
case of the transcription regulation network.

  \begin{table}[t]
\caption{Estimates for the simple gene expression model using
equidistant time series.\label{tab:geneexp_eq}}
\begin{center}\scalebox{0.9}{
\begin{tabular*}{1.0\textwidth}{@{\extracolsep{\fill}} l l l r l l l l }
\hline\hline
$\Delta t~(R)$ \qquad & $\sigma$ & Method & Time &
\multicolumn{4}{l}{\quad Average (standard deviation) of parameter estimates}\\
                          &        &          & & $c_1=0.027$  & $c_2=0.1667$  &
$c_3=0.4$ & $\quad\sigma$ \\ \hline
1.0~(300)           & 0.1  & \algR  & --\quad & 0.0268(0.0061) & 0.1523(0.0424) & 0.3741(0.0557) & 0.1012(0.0031) \\
						  &        & \algS & -- & 0.0229(0.0041) & 0.1573(0.0691) & 0.4594(0.1923) & \quad --      \\
                          &        & \algV  & 29.4 & 0.0297(0.0051) & 0.1777(0.0361) & 0.3974(0.0502) & 0.1028(0.0612) \\
                          &        & \algP  & 2.2 & 0.0300(0.0124) & 0.1629(0.0867) & 0.3892(0.0972) & 0.1010(0.0792) \\ \hline
                          & 1.0  & \algR  & --\quad & 0.0257(0.0054) & 0.1409(0.0402) & 0.3461(0.0630) & 1.0025(0.0504) \\
						  &        & \algS & -- & 0.0295(0.0102) & 0.1321(0.0787) & 0.3842(0.2140)  & \quad --      \\
                          &        & \algV  & 8.3 & 0.0278(0.0047) & 0.1868(0.0339) & 0.3946(0.0419) & 0.9976(0.0476)  \\
                          &        & \algP  & 1.8 & 0.0278(0.0041) & 0.1810(0.0294) & 0.3938(0.0315) & 0.9938(0.0465) \\ \hline
                          & 3.0  & \algR  & --\quad & 0.0250(0.0065) & 0.1140(0.0337) & 0.3160(0.0674) & 3.0292(0.1393) \\
                          &        & \algS  & --\quad & \quad -- & \quad -- & \quad -- & \quad -- \\
                          &        & \algV  & 11.1 & 0.0285(0.0043) &  0.1755(0.0346) & 0.3938(0.0508) & 2.9913(0.0733)         \\
                          &        & \algP  & 1.7  & 0.0275(0.0086) & 0.1972(0.0902) & 0.3894(0.0722) & 3.0779(0.0887) \\ \hline
10.0~(30)           & 0.1  & \algR  & --\quad & \quad -- & \quad -- &  \quad -- &  \quad -- \\
						  &        & \algS & -- & \quad  -- & \quad -- &  \quad -- &  \quad -- \\
                          &        & \algV  & 40.9 & 0.0273(0.0069) & 0.1788(0.04786) & 0.3931(0.0599) & 0.1086(0.0630) \\
                          &        & \algP  & 5.2 & 0.0277(0.0080) & 0.1782(0.0517) & 0.4057(0.0678) & 0.1234(0.0523) \\ \hline
                          & 1.0  & \algR  &  --\quad & \quad -- & \quad -- &  \quad -- &  \quad -- \\
                          &        & \algS & --\quad & \quad -- & \quad -- &  \quad -- &  \quad -- \\
                          &        & \algV  & 10.2 & 0.0283(0.0070) & 0.1787(0.0523) & 0.4018(0.0681) & 0.9898(0.0829) \\
                          &        & \algP  & 3.5 & 0.0243(0.0057) & 0.1665(0.0400) & 0.4031(0.0638) & 1.0329(0.0859) \\ \hline
                          & 3.0  & \algR  & --\quad & \quad -- & \quad -- &  \quad -- &  \quad -- \\
                          &        & \algS &  --\quad & \quad -- & \quad -- &  \quad -- &  \quad -- \\
                          &        & \algV  & 12.3 & 0.0300(0.0110) & 0.1960(0.0788) & 0.4025(0.0689) & 2.9402(0.1304) \\
                          &        & \algP  & 4.2 & 0.0210(0.0054) & 0.1511(0.0534) &  0.4042(0.0616) & 3.0629(0.2249) \\ \hline           
\end{tabular*}}
\end{center}\vspace{-4ex}
 \end{table}%
\subsection{Equidistant time series}\vspace{-0.8ex}
 In the equidistant case, the length of the observation intervals is  
 $\Delta t=t_{\ell} - t_{\ell-1}$ for all $\ell\in\{1,\ldots,R\}$.
In Table~\ref{tab:geneexp_eq} and~\ref{tab:goutsias_eq} we list the results
given in~\cite{Timmer} as well as the results of our methods.   Reinker
et al. do not evaluate the AML method for larger intervals than $\Delta t=1$
because, as we will discuss in Section~\ref{sec:related}, 
 the approximation error of the AML method 
becomes huge in that case.
Also, the SVDL method performs poor if $\sigma$ is
large since it does not include measurement errors in the 
likelihood. Therefore, no results for $\sigma>1.0$ are provided 
in~\cite{Timmer} for SVDL.
In the first three columns we list $\Delta t$, the number $R$ of observation
points and the true standard deviation $\sigma$ of the error terms.
In column ``Time'' we compare the average running time (in seconds) 
 of one parameter
estimation  (out of 100) for   \algV~and \algP, i.e., the average running time 
of the maximization of the likelihood based on $K=5$ observation sequences.
It is not meaningful to compare the running times with those in~\cite{Timmer}
since different optimization methods are used and experiments were run on
different machines.
 Finally, we list estimation results for all four methods (if available). We
give the true parameters in the column headings and list the average of 100
estimations and the   standard deviation of the estimates (in brackets).

For the simple gene expression (Table~\ref{tab:geneexp_eq}) 
and $\Delta t=1.0$, we find that 
SLA and PLA have a similar accuracy for the estimation of $\sigma$ 
but are consistently more accurate than AML and SVDL 
for estimating the rate constants.
If $\sigma=0.1$, then the total absolute error for the 
estimation of $\bfc$ is 0.041, 0.073, 0.016, 0.018 
for   AML, SVDL, SLA, PLA, respectively. 
For $\sigma=1.0$ we have total absolute errors of 
0.081, 0.053, 0.026, 0.021 
for   AML, SVDL, SLA, PLA.
Finally, for $\sigma=3.0$, AML has a total   error of
0.139 while the error for SLA and  PLA is 0.017 and 0.041.
For $\Delta t=10$, the results of the SLA and  PLA method 
are accurate even though only 30 observation points are given.
Since PLA gives a much coarser approximation, its running time 
is always shorter (about three to ten times shorter).
If $\sigma$ is large, SLA gives more 
accurate results than PLA.   

In Table~\ref{tab:goutsias_eq} we compare results of the  transcription
regulation for   $\sigma=0$.
 Note that, for this example, Reinker et al. only give results
 for  the SVDL method with $\Delta t\le 1.0$ and $\sigma=0$.
Here, we compare   results  for   $\Delta t= 1.0$ since in this case the SVDL
method performs best compared to smaller values of $\Delta t$.
The SLA and  PLA method consistently  perform better than the SVDL method since 
they approximate the likelihood more accurately. 
If $\sigma=0$, then the accuracy of  SLA and  PLA is the same 
(up to the fifth digit). 
Therefore the results of SLA and PLA  are combined in
Table~\ref{tab:goutsias_eq}. The running time of SLA is, however, much slower
since it does not reuse the entries of the transition probability matrix $T$. 
For $\Delta t=1.0$, one parameter estimation based on $K=5$ observations 
takes about 30 minutes for SLA and 
only about 2.4 minutes for PLA. For $\Delta t=10.0$  we have 
 running times about 5 hours(SLA) and 27 minutes (PLA). 
  As for the gene expression example, we expect for larger values of $\sigma$ 
the results of SLA to be more accurate than those of PLA. 
 \subsection{Non-equidistant time series}\vspace{-0.8ex}
Finally, we consider non-equidistant time series, which can only
be handled by the SLA method.
During the Monte-Carlo simulation, we generate non-equidistant time series by
iteratively choosing $t_{\ell+1} = t_{\ell} +
\mathcal{U}(0,5)$, where $\mathcal{U}(0,5)$ is a random number that is
uniformly distributed on  $(0,5)$ and $t_0=0$.  Note that the 
intervals are not only different within an observation sequence
 but also for different $k$, i.e., the   times $t_1,\ldots,$ $t_R$
depend on the number $k$ of the corresponding sequence.
We consider the  transcription regulation model 
with $\sigma=1.0$ and $K=5$ as this is our most complex example.
Note that, since the accuracy of the estimation decreases as $\sigma$ 
increases, we cannot expect a similar accuracy as in Table~\ref{tab:goutsias_eq}. 
For a time horizon of $t=500$ the average number of
observation points per sequence is $R=500/2.5=200$.
The estimates computed by SLA are
 $c_1^*=0.0384(0.0343)$,  $c_2^*=0.0010(0.0001)$, $c_3^*=0.0642(0.0249)$, $c_4^*=0.0044(0.0047)$, $c_5^*=0.0273(0.0073)$, $c_6^*=0.5498(0.1992)$, $c_7^*=0.0890(0.0154)$,
 $c_8^*=0.5586(0.0716)$, and $\sigma^*=0.9510(0.0211)$, 
where we averaged over 100 repeated estimations and give the 
standard deviation  in brackets.
Recall that the true  constants are $c_1 = 0.043$, 
$c_2=0.0007,$ $c_3=0.0715$, $c_4=0.00395,$ $c_5=0.02$, $c_6=0.4791$, 
$c_7=0.083$, and $c_8=0.5$.
 The average running time of one estimation was 
  19 minutes. 
\begin{table}[t]
\caption{Estimates for the transcription regulation model using
equidistant time series.\label{tab:goutsias_eq}}
\begin{center}
\scalebox{0.9}{
\begin{tabular*}{1.0\textwidth}{@{\extracolsep{\fill}} l l l l l l l}
\hline\hline
$\Delta t~(R)$ \qquad\quad & Method   & \multicolumn{4}{l}{Average (standard
deviation) of parameter estimates}\\
                          &             & $c_1=0.043$  & $c_2=0.0007$  &
$c_3=0.0715$  & $c_4=0.00395$   \\ \hline
1.0~(500)           & \algS  & 0.0477(0.0155 ) & 0.0006(0.0004) &
0.0645(0.0190) & 0.0110(0.0195)    \\
                          & \algP/\algV  & 0.0447(0.0036) &
0.0007(0.0001) &
0.0677(0.0115) & 0.0034(0.0014)   \\
\hline                   
10.0~(50)            & \algP/\algV & 0.0417(0.0069) & 0.0005(0.0002) &
0.0680(0.0075) & 0.0038(0.0026) \\[2ex]
\hline\hline
$\Delta t~(R)$ \qquad & Method    & \multicolumn{4}{l}{Average (standard
deviation) of parameter estimates}\\
                          &                 & $c_5=0.02$  & $c_6=0.4791$  &
$c_7=0.083$  & $c_8=0.5$ \\ \hline
1.0~(500)           & \algS     & 0.0159(0.0107) & 0.2646(0.0761) &
0.0149(0.0143) & 0.0615(0.0332) \\
                          & \algP/\algV     & 0.0193(0.0008) & 0.4592(0.0169) &
0.0848(0.0024) & 0.5140(0.0166) \\
\hline                   
10.0~(50)                                 & \algP/\algV     & 0.0188(0.0039) &
0.4359(0.0822) & 0.0836(0.0016) & 0.4892(0.0164) \\
\hline                   
\end{tabular*}
}
\end{center}
\vspace{-4ex}\end{table}%

\section{Related work} \label{sec:related}
 In the context of stochastic chemical kinetics, 
parameter inference methods are either based on 
Bayesian inference~\cite{Wilkinson2,Stumpf,Wilkinson}
or maximum likelihood estimation~\cite{Timmer,Uz2010478,Tian2007}. 
The advantage of the latter method is that the corresponding 
estimators are, in a sense, the most 
informative estimates of   unknown parameters~\cite{Higgins}
and have desirable mathematical properties such as
    unbiasedness,
   efficiency, and    normality~\cite{citeulike:821121}. 
On the other hand, the  computational complexity of 
maximum likelihood estimation is high. If an analytic solution 
of~\eqref{eq:MLEestimator} is not possible,
then, as a part of the nonlinear optimization problem, 
the   likelihood and its derivatives have to be calculated.
Monte-Carlo simulation has been used to estimate the 
likelihood~\cite{Tian2007,Uz2010478}. During the 
repeated random sampling it is difficult to explore those parts of the state
space  that are unlikely under the current rate parameters.
Thus, especially if the rates are very different from the true parameters,
many simulation runs are necessary to calculate an accurate 
approximation of the likelihood.
To the best of our knowledge, Reinker et al. provide the first 
 maximum likelihood estimation
that is not based on Monte-Carlo simulation but  calculates the
likelihood numerically~\cite{Timmer}.
They propose the AML method during which  the 
matrices $P_\ell$ are approximated. In order to keep the computational effort
low, they allow at most two jumps of the Markov process during 
$[t_\ell,t_{\ell+1})$.  
Moreover, they ignore
all states for which $|O_i(t_\ell)-x_{i\ell}|$ is greater than 
$3\sqrt{\sigma}$. This has the disadvantage that $\caL$ is zero 
(and its derivative) if the values for the rate constants are
far off the true values. If $\caL$ is zero, then  the derivatives provide no
information about how the rate constants have to be altered in order 
to increase the likelihood. Thus, initially very good estimates for the rate 
constants must be known to apply this kind of truncation.
On the other hand,  the method that we propose neglects only   
insignificant terms of the likelihood. For this reason the 
likelihood and its derivatives  do not become zero during the computation
  and it is always possible to follow the gradient in order
to obtain higher likelihoods.
Another disadvantage of the AML method is that,
if the observation intervals are longer, the likelihood may not be
 approximated accurately since the assumption that only two
reactions occur within an observation interval is not valid.
 Extending the AML approach to more than two steps would result in huge 
space requirements and perform slow since the state space is explored in a
breath-first search manner and too many states would be considered even though
their contribution to the likelihood is very small.  
In our approach   we allow an arbitrary number of reactions during 
$[t_\ell,t_{\ell+1})$\footnote{During one step of our numerical 
integration, we assume that only four reactions are possible. The time 
step $h$ of the numerical integration does, however, not depend on the 
$[t_\ell,t_{\ell+1})$ but is dynamically chosen in such a way that performing
more than four steps is very unlikely.}.
Therefore, our method is not restricted to reaction networks where 
the speed of all reactions is at most of the same time scale as 
the observation intervals. 
 The second approach proposed by Reinker et al., called SVDL method,
is based on the assumption that the propensities $\alpha_j$ stay constant 
during $[t_\ell,t_{\ell+1})$. Again, this assumption only applies to small
observation intervals. 
Moreover, the SVDL method does not take into 
account measurement errors and is thus only appropriate if $\sigma$ 
is very small.
 Further differences between the approach of Reinker et al. and our approach 
are that we use a global optimization technique (MATLAB's global search) while 
Reinker et al. use a local solver, namely the quasi-Newton method.
Finally, the approach in~\cite{Timmer} requires   observations at
equidistant time instances, which is not necessary for the SLA method.

\section{Conclusion}
Parameter inference for stochastic models of cellular processes 
demands huge computational resources. We proposed two numerical 
 methods, called SLA and PLA, that approximate maximum likelihood 
estimators for a given set of observations.
Both methods do not make any assumptions about the number 
of reactions that occur within an observation 
interval. The SLA method allows for an estimation based on arbitrarily 
spaced intervals while the
PLA method requires equidistant intervals.

Many reaction networks involve both small populations 
and large populations. In this case stochastic hybrid models 
are most appropriate since 
they combine the advantages of deterministic and stochastic 
representations. We plan to extend our algorithms to the 
stochastic hybrid setting proposed in~\cite{CMSB10} to allow inference for 
more complex networks.  
Further future work also includes more rigorous truncations for
the SLA method and the parallelization of  the algorithm.

\bibliographystyle{plain}
\bibliography{inference}  

\end{document}